\documentclass[12pt,a4paper]{article}

\title{Dynamical generation of the constituent mass in expanding plasma}

\author{{I.N. Mishustin$^{1,2}$,  O. Scavenius$^{1}$} \\
{\it $^{1}$The Niels Bohr Institute, University of Copenhagen,}\\
{\it Blegdamsvej 17. DK-2100 Copenhagen \O , Denmark and} \\ 
{\it $^{2}$The Kurchatov Institute, Russian Scientific centre,}\\
{\it Moscow, 123182 Russia,}}

\begin{document}

\maketitle

\begin{abstract}
We investigate dynamics of the chiral transition in expanding quark-antiquark plasma produced in an ultra-relativistic heavy ion collision. The chiral symmetry break-down and dynamical generation of the constituent quark mass are studied within the linear sigma model and Nambu\---Jona-Lasinio model. Time dependence of the quark and antiquark densities is obtained from the scaling solution of the relativistic Vlasov equation. Fast initial growth and strong oscillations of the constituent quark mass are found in the linear sigma model as well as in the NJL model, when derivative terms are taken into account. 
\end{abstract}

 \\


relativistic heavy ion collisions, scaling expansion.\\
 \\

{\em Introduction.--} It is commonly believed that colour deconfinement and chiral symmetry restoration take place at early stages of ultra-relativistic heavy-ion collisions. At intermediate stages of the reaction the quark-gluon plasma may be formed and evolve through the states close to thermodynamical equilibrium. However, at later stages of the expansion the transition to the hadronic phase with broken chiral symmetry should take place. The break-down of chiral symmetry will possibly lead to such interesting phenomena as formation of disoriented chiral condensates (DCCs) and classical pion fields as well as clustering of quarks and antiquarks. These phenomena were studied recently in many publications [1-14], using QCD motivated effective models, such as the linear and non-linear sigma models and the Nambu\---Jona-Lasinio (NJL) model. Of course, these models have some significant shortcomings, e.g. they do not possess colour confinement and the NJL-model is non-renormalizable. The key point is, however, that these models obey the same chiral symmetry as the QCD Lagrangian.

In most applications of the sigma model the quark degrees of freedom are disregarded (see e.g. [2-9,14]). The inclusion of quarks \cite{mishu,mismocs}  makes it possible to study the hadronization process, in particular, the dynamical generation of the constituent quark mass. In this letter, we consider the late stages of the plasma evolution, such that collisions between quarks and antiquarks are not frequent enough to maintain thermodynamical equilibrium. We assume that the expansion is spherical at this stage. We use the linear sigma model and the NJL model to describe the interaction of quarks with background chiral fields.

{\em Linear sigma model.--} The Lagrangian density of the linear sigma model with quark degrees of freedom reads
\begin{equation}
{\cal L}=\overline{q}[i\gamma ^{\mu}\partial _{\mu}-g(\sigma +i\gamma _{5}\vec{\tau} \cdot \vec{\pi} )]q
+ \frac{1}{2}(\partial _{\mu}\sigma \partial ^{\mu}\sigma + \partial _{\mu}\vec{\pi} \partial ^{\mu}\vec{\pi} )-U(\sigma ,\vec{\pi} ), 
\label{sigma}
\end{equation}
where
\begin{equation}
U(\sigma ,\vec{\pi} )=\frac{\lambda ^{2}}{4}(\sigma ^{2}+\vec{\pi} ^{2} -{\it v}^{2})^{2}-H\sigma
\end{equation}
is the Mexican Hat potential. Here $q$ is the light quark field $q=(u,d)$. The
scalar field $\sigma$ and the pion field $\vec{\pi} =(\pi _{1},\pi _{2},\pi _{3})$ form together a chiral field $\Phi =(\sigma,\vec{\pi})$. This Lagrangian is approximately invariant under chiral $SU_{L}(2) \otimes SU_{R}(2)$ transformations if the explicit symmetry breaking term $H\sigma $ is small. The parameters of the Lagrangian are chosen so that the chiral symmetry is spontaneously broken in the vacuum and the expectation values of the meson fields are $\langle\sigma\rangle ={\it f}_{\pi}$ and $\langle\vec{\pi}\rangle =0$, where ${\it f}_{\pi}=93$ MeV is the pion decay constant. The constant $H$ is fixed by the PCAC relation that gives $H=f_{\pi}m_{\pi}^{2}$, where $m_{\pi}=138$ MeV is the pion mass. Then one finds $v^{2}=f^{2}_{\pi}-\frac{m^{2}_{\pi}}{\lambda ^{2}}$. The $\lambda ^{2}$ is determined by the sigma mass, ${m_{\sigma}}^{2}=2\lambda ^{2}f^{2}_{\pi}+m^{2}_{\pi}$, which we set to 600 MeV yielding $\lambda ^{2} \approx 20$. The coupling constant g is fixed by the requirement that the constituent quark mass in vacuum, $m_{vac}=gf_{\pi}$, is about $1/3$ of the nucleon mass, that gives
$g \approx 3.3$.  With these parameters a chiral phase transition is predicted at $T_{c} \approx 132$ MeV \cite{mismocs}.

Below we adopt the mean field approximation, considering $\sigma$ and $\vec{\pi}$ as classical fields. The variation of  ${\cal L}$ with respect to $\sigma $ and $\vec{\pi} $ yields the equations of motion
\begin{displaymath}
\partial _{\mu}\partial ^{\mu}\sigma (x)+\lambda ^{2}[\sigma ^{2} (x)+\vec{\pi} ^{2} (x)-v^{2}]\sigma (x)-H=-g\rho _{s}(x),
\end{displaymath}
\begin{equation}
\partial _{\mu}\partial ^{\mu}\vec{\pi} (x)+\lambda ^{2}[\sigma ^{2} (x)+\vec{\pi}^{2} (x)-v^{2}]\vec{\pi} (x)=-g\vec{\rho}_{p}(x),
\label{sigmab}
\end{equation}     
where $\rho _{s}=\langle\overline{q}q\rangle$ and $\vec{\rho}_{p}=i\langle\overline{q}\gamma _{5}\vec{\tau} q\rangle$ are respectively the scalar and pseudoscalar densities of valence quarks. They are determined by the interaction of quarks with meson fields. As shown in \cite{mishu} these densities can be expressed as $\rho _{s}(x)=g\sigma (x)a(x)$ and $\vec{\rho}_{p}=g\vec{\pi} (x)a(x)$, with scaling function 
\begin{equation}
a(x)=\frac{\nu _{q}}{(2\pi )^{3}}\int \frac{d^{3}p}{\sqrt{{\bf p}^{2}+m^{2}(x)}}[n_{q}(x,{\bf p})+n_{\overline{q}}(x,{\bf p})]
\label{af}
\end{equation}
Here $\nu_{q}=2\times2\times3=12$ is the degeneracy factor of quarks and $m(x)$ is the constituent quark mass which is determined self-consistently through the meson fields,
\begin{equation}
m^{2}(x)=g^{2}[\sigma ^{2}(x)+\vec{\pi}^{2}(x)].
\end{equation} 
In eq. (\ref{af}) $n_{q}$ and $n_{\overline{q}}$ are the quark and antiquark occupation numbers which in thermal equilibrium are given by the Fermi-Dirac distributions. 

{\em Nambu\---Jona-Lasinio model.--} Now we turn to the Nambu\---Jona-Lasinio model which is widely used now for describing hadron properties and the chiral transition. The Lagrangian as introduced by Nambu and Jona-Lasinio \cite{njl} in
1961 is given by:
\begin{equation}
{\cal L}=\overline{q}(i\gamma^{\mu} \partial_{\mu} -m_{0})q+\frac{G}{2} \left[(\overline{q} q)^{2}+(\overline{q} i\gamma_{5} \vec{\tau} q)^{2}\right],
\label{njl}
\end{equation}
where $q$ stands for the quark field and $m_{0}$ is the small current quark mass. At vanishing $m_{0}$, the NJL Lagrangian is invariant under the chiral $ SU_{L} (2) \otimes  SU_{R} (2) $ transformations.  The coupling constant $G$ has dimension $($energy$)^{-2}$, that is why the theory is non-renormalizable. Therefore, a cutoff momentum  $\Lambda$ is introduced to regularize divergent integrals. It defines an upper energy limit for this effective theory. Free parameters of the model are fixed to reproduce correctly the vacuum values of the pion decay constant, pion mass and the constituent quark mass. When the current quark mass $m_{0}$ is set to zero, the remaining parameters are $G=11$ GeV$^{-2}$ and $\Lambda =653$ MeV \cite{klevansky}. With these parameters the chiral transition occurs at critical temperature $T_{c}=190$ MeV \cite{klevansky} which is significantly higher than in the sigma model.

In the mean-field approximation the Lagrangian (\ref{njl}) is represented in a linearized form \cite{weise} where the constituent quark mass is expressed as
$\hat{m}(x)=m_{s}(x)+i\gamma_{5}\vec{\tau} \cdot \vec{m}_{p}(x)$. By definition
\begin{equation}
m_{s}(x)=-G{\langle}{\hskip -2pt}{\langle} \overline{q}(x)q(x){\rangle}{\hskip -2pt}{\rangle}\mbox{, }\vec{m}_{p}(x)=-G{\langle}{\hskip -2pt}{\langle} i\overline{q}(x)\gamma_{5}\vec{\tau}q(x){\rangle}{\hskip -2pt}{\rangle} ,
\end{equation}
where ${\langle}{\hskip -2pt}{\langle} ...{\rangle}{\hskip -2pt}{\rangle}$ means averaging over the exact many-body state including the Dirac sea. The right hand sides of these expressions can be expressed through the quark Green's function $S(x,y)$ which is determined  self-consistently with the exact mass $\hat{m}(x)$. In the case of slow varying mass the scalar and pseudoscalar densities can be parametrized in terms of quark and antiquark occupation numbers as $\tilde{\rho}_{s}(x)=m_{s}(x)\tilde{a}(x)$ and $\vec{\tilde{\rho}}_{s}(x)=\vec{m}_{p}(x)\tilde{a}(x)$ where
\begin{equation}
\tilde{a}(x)=\frac{\nu_{q}}{(2\pi)^{3}}\int\frac{d^{3}p}{\sqrt{{\bf p}^{2}+m^{2}(x)}}[n_{q}(x,{\bf p})+n_{\overline{q}}(x,{\bf p})-1].
\label{atf}
\end{equation}
Here the term with $-1$ corresponds to the contribution from the Dirac sea and the rest comes from valence quarks and antiquarks (compare with eq. (\ref{af}))

In the case of rapidly changing $m(x)$ one should solve the exact consistency relations in terms of $S(x,y)$. For this purpose the derivative expansion method was proposed by Eguchi and Sugawara \cite{eghusu}. It consists in expanding $S(x,y)$ around the Green's function $S_{0}(x,y)$ for a homogeneous state and retaining only divergent terms in the momentum-space integrals.
In contrast to \cite{eghusu}, where only the vacuum contribution was considered, here we expand around the state characterized by the scalar and pseudoscalar densities $\tilde{\rho}_{s}(x)$ and $\tilde{\vec{\rho}}_{p}(x)$. Then, with minor modifications, one can repeat all steps of the derivation and arrive at the modified gap-equation:
\begin{equation}
    \partial_{\mu}\partial^{\mu}\hat{m}(x)-2G^{2}[\tilde{\rho}^{2}_{s}(x)+\tilde{\vec{\rho}}^{2}_{p}(x)]\hat{m}(x)+2|\hat{m}(x)|^{2}\hat{m}(x)=0.
\label{ggap}
\end{equation}
Here $|\hat{m}(x)|^{2}=m^{2}(x)=m^{2}_{s}(x)+\vec{m}^{2}_{p}(x)$ and $\tilde{\rho}^{2}_{s}(x)+\tilde{\vec{\rho}}^{2}_{p}(x)=m^{2}(x)\tilde{a}^{2}(x)$, where $\tilde{a}^{2}(x)$ is given by eq.
 (\ref{atf}). In the limit of a homogeneous system ($\partial_{\mu}=0$) this equation is reduced to 
$G^{2}\tilde{a}^{2}=1$ which is equivalent to the famous gap equation used in most applications of the NJL model. Eq. (\ref{ggap}) can be easily generalized to include the explicit symmetry breaking terms. By expressing $m_{0}$ in terms of $m^{2}_{\pi}$ they can be represented in the form analogous to eq. (\ref{sigmab}). 

One can easily see that in vacuum eq. (\ref{ggap}) is equivalent to the linear sigma model discussed above. Indeed, in this case $\tilde{\vec{\rho}}_{p}=0$ and the gap equation has the form $m_{vac}=-G\tilde{\rho}_{s}$, where $m_{vac}$ is the constituent quark mass in vacuum, which is written above as $m_{vac}=gf_{\pi}$. Then eq. (\ref{ggap}) coincides with eqs. (\ref{sigmab}) if one chooses $\lambda ^{2}=2g^{2}$. (In fact this is very close to our choice of parameters). With this choice $m_{\sigma}=2m_{vac}$ which is quite natural for the composite $\sigma$-meson of the NJL model. However, we cannot prove equivalence of the two models when valence quarks are present. This is because the vacuum contribution to scalar density is strongly affected by the valence quarks and antiquarks. This effect is simply ignored in the sigma model. In contrast, the Dirac sea and valence quarks are treated self-consistently in the NJL model. In this respect it represents a more fundamental theory.   

{\em Scaling expansion.--} The evolution of a system out of equilibrium is governed by transport equations. For our purpose we use the relativistic Vlasov equation consistent with both the linear sigma model \cite{mishu} and the NJL model \cite{klevansky,oth,wil}: 
\begin{equation}
\left[p_{\mu}\frac{\partial}{\partial x_{\mu}}+\frac{1}{2}\frac{\partial m^{2}(x)}{\partial x_{\mu}} \cdot \frac{\partial}{\partial p_{\mu}}\right]{\it f}(x,p)=0.
\label{vlasov}
\end{equation} 
Here $f(x,p)$ is the scalar part of the fermion distribution function. In general $f(x,p)$ depends on 4-momentum $p$ and 4-coordinate $x$. On the mass shell $p^{\mu}p_{\mu}=m^{2}(x)$ and $f(x,p)$ goes over into the sum of quark and antiquark occupation numbers as written in eqs. (\ref{af}) and (\ref{atf}). The vanishing collision integral in the r.h.s. of eq. (\ref{vlasov}) corresponds to a situation when collisions between quarks and antiquarks cease i.e. after thermal freeze-out. This might be a reasonable approximation for later stages of the expansion when the temperature drops below $140$ MeV and the mean free path of quarks and antiquarks becomes long \cite{klevansky}. 

It is not trivial to find a solution of the Vlasov equation in a general case of ($3+1$) dimensional evolution. But in this letter we consider a simplified case of the spherical and homogeneous (Hubble-like) expansion \cite{bjork}. It is characterised by the collective $4$-velocity $u^{\mu}=x^{\mu} / \tau$, where $\tau =\sqrt{t^{2}-{\bf r}^{2}}$ is the proper time and ${\bf r}=(x,y,z)$. Assuming that $m(x)$ depends on one variable $\tau$ only, one can show \cite{mishu} that a general solution of the Vlasov equation (\ref{vlasov}) is any function of the scaling variable $\tau |{\bf p}|$. This function can be specified under the assumption that at the time of freeze-out, $\tau=\tau _{0}$, the quark and anti-quark occupation numbers are given by the equilibrium Fermi-Dirac distributions. Then, at $\tau >\tau _{0}$ the quark and antiquark occupation numbers can be obtained as: 
\begin{equation}
n_{q}(\tau,{\bf p})=\left[ \exp\left(\frac{\sqrt{(\frac{\tau }{\tau _{0}}{\bf p})^{2}+m^{2}(\tau _{0})}-\mu_{0}}{T_{0}}\right)+1 \right]^{-1},\mbox{ } n_{\overline{q}}(\tau,{\bf p})=n_{q}(\tau,{\bf p}, \mu_{0}\rightarrow -\mu_{0}), 
\label{occu}
\end{equation}
where $m(\tau _{0})$, $T_{0}=T(\tau_{0})$ and $\mu_{0}=\mu (\tau_{0})$ are the constituent mass, temperature and chemical potential at $\tau =\tau _{0}$. Below we set the chemical potential to zero, $\mu _{0}=0$, so that $n_{q}=n_{\overline{q}}$. The initial state for the scaling expansion is chosen at the respective critical temperatures for both models, i.e $T_{0}=T_{c}$. 

In the case of homogeneous spherical expansion the d'Alembertian is reduced to $\partial_{\mu}\partial^{\mu}\longrightarrow\frac{d^{2}}{d\tau^{2}}+\frac{3}{\tau}\frac{d}{d\tau}$. Now equations (\ref{sigmab}) and (\ref{ggap}) are ordinary differential equations which can be solved numerically by standard methods \cite{numres}. In numerical calculations the initial values of fields and their first derivatives were selected within the range determined by the root-mean-square values  of corresponding thermal fluctuations at $\tau =\tau_{0}$. At times $\tau >\tau _{0}$ the field fluctuations do not stay in thermal equilibrium but rather ``propagate'' according to equations of motion.

In the quench scenario, introduced in ref. \cite{rajg},  one considers thermal equilibrium initially and then lets the fields evolve out of equilibrium in the zero-temperature potential $U(\sigma ,\vec{\pi} )$ (\ref{sigma}). The fields roll down from the unstable local maximum of the Mexican Hat potential at $\sigma \approx 0, \vec{\pi}\approx 0$ towards the stable minimum at  $\sigma ={\it f}_{\pi }$ and $\vec{\pi}=0$ and oscillate around these values. This quench scenario corresponds to a sudden removal of the heatbath which initially provides thermal and dynamical equilibrium. As pointed out in \cite{gm}, a more realistic scenario is when the potential changes gradually during the evolution and finally approaches $U(\sigma ,\vec{\pi} )$. Gavin and M{\"u}ller \cite{gm} and recently Randrup \cite{randrup} considered a model where the fields evolve in an effective potential generated by mesonic quasi-particles treated in the Hartree approximation (hot annealing). Our picture is somewhat similar, but instead of mesonic quasi-particles we consider quarks and antiquarks. However, there exists an important difference between \cite{gm,randrup} and the present approach. In refs. \cite{gm,randrup} the quasi-particles are assumed in thermal equilibrium throughout the whole evolution, while in our approach this assumption refers to the initial state only. At later times quarks and antiquarks follow the scaling solution of the Vlasov equation which differs from the thermal equilibrium distribution.

{\em Numerical results,--} Fig 1 shows the effective potential for different proper times as predicted by our model. One clearly sees that the potential evolves quite slowly and approaches the zero-temperature potential $U(\sigma, \vec{\pi})$ only at about $5\tau_{0}$. The asymmetry of the potential is due to the symmetry-breaking term $H\sigma$ in the Lagrangian. Note that the effective potential depends on the initial conditions for the fields through the constituent quark mass appearing in (\ref{occu}). The behaviour of the effective potential for the NJL model is qualitively similar.

Now let us discuss numerical solutions  of the equations of motion for scaling expansion. In the case of the sigma model we consider an initial temperature  $T_{0}=132$ MeV and initial proper time $\tau _{0}=7$ fm/c  \cite{mishu,mismocs} .  At this temperature the root-mean-square fluctuations of fields and their derivatives are $38$ MeV and $35$ MeV/fm, respectively. Fig 2 displays the meson fields together with the constituent quark mass as functions of proper time. One observes strong oscillations of both the constituent quark mass and the fields. The complicated shape of the curves for the $\sigma$-field and quark mass is due to the coupling of oscillations in different directions in isospin space. The constituent mass and the $\sigma$-field first cross their equilibrium values already at about $1.3\tau_{0}$, although the effective potential at this time is still far away from the assymptotic shape. Notice the strong long-wavelength oscillations of the $\pi$-field that may lead to rather big DCC domains \cite{gm}. 

It is instructive to compare three different scenarios: quench \cite{rajg}, annealing \cite{gm} and the collisionless expansion as discussed above. Fig. 3 shows the $\sigma$-field evolution for these three cases. The initial conditions are all around zero. In the quench scenario we see a large overshoot in the $\sigma$-field of about $33$ MeV compared to  the equilibrium value of $f_{\pi}$, then it oscillates strongly about the equilibrium value. This is a direct consequence of the zero-temperature potential used in this case. The fields just ``glide'' down from the top of the Mexican Hat to the minimum and then ``climb'' up again due to the large kinetic energy. In the annealing scenario we clearly observe the effect of the gradual change of the effective potential. The fields go ``softer'' towards equilibrium. In our scenario the effect of the changing effective potential (see Fig 1) is quite similar, except for  stronger oscillations and slightly faster equilibration. 

In fig. 4 we compare the proper time evolution of the constituent quark mass calculated
within the linear sigma-model and the NJL model. The initial state for the NJL model is chosen at $T_{0}=190$ MeV corresponding to $\tau _{0}=5$ fm/c. Curve a) is calculated by combining eq. (\ref{occu}) with the standard gap equation for the NJL model, $G\tilde{a}(x)=-1$. Since in this lowest order approximation the gap equation does not contain any derivative terms, the constituent quark mass evolves smoothly to the vacuum value $m_{vac}$. One can say that it just follows the minimum of the evolving effective potential. In contrast, when eq. (\ref{occu}) is combined with the generalized gap equation (\ref{ggap}) the character of solution changes drastically (curve (b)). First of all, one sees much faster rise of the constituent mass right after the beginning of the non-equilibrium stage. Second, the approach to the asymptotic value goes through strong oscillations, i.e. very similar to the sigma model (curve (c)). In the NJL case the oscillations have slightly longer period and smaller amplitude than in the sigma model. This is partly related to the difference in the critical temperatures and the corresponding initial times. As already mentioned above, this difference comes from the different treatment of the Dirac sea and valence quarks in these two models. 
     
{\em Summary.--} We applied two chiral models with quark degrees of freedom, i.e. the linear sigma model and the NJL model, to study dynamics of the chiral transition in expanding quark-antiquark plasma. A generalized gap equation with derivative terms was used for the NJL model. The calculations were carried out for a spherical homogeneous system and comparison of three different scenarios, quench, hot annealing and collisionless expansion, was made. Qualitively similar oscillatory behaviour of the constituent quark mass was found in all cases. We conclude that strong space-time fluctuations of the chiral fields are characteristic for a rapid chiral transition and they can be used for its experimental detection. In the future we are planning to study DCC domain formation and clustering of quarks as well as dissipative effects and soft pion production in the course of the chiral transition
 \\
 \\
{\bf {\it Acknowledgement.--}} The authors thank R.F. Bedaque, J. Bjorken, D. Boyanovsky, T.S. Bir{\' o}, J.P. Bondorf, L.P. Csernai, A.D. Jackson, S. Klevansky, {\' A}. M{\' o}csy, J. Randrup and L.M. Satarov for stimulating discussions. One of us (I.N.M.) thanks the organizers and participants of the Workshop 
on Disoriented chiral Condensates at ECT* in Trento (October 1996) for the fruitful atmosphere. This work was supported in part by EU-INTAS grand N\b{o} 94-3405. We are grateful to the Carlsberg Foundation  and to the Rosenfeld Fund for financial support.

{\bf figure captions}\\
 \\
 \\
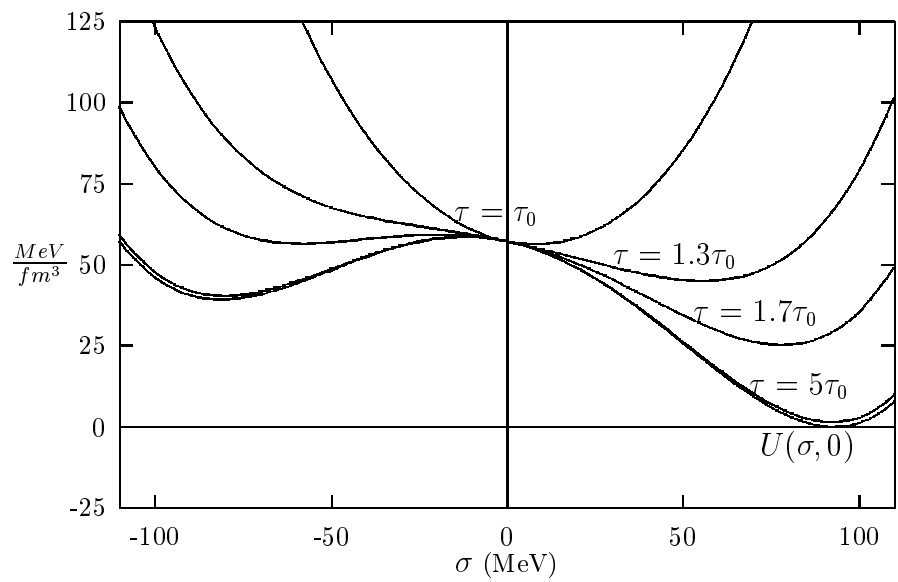
Figure 1: Effective potential as a function of $\sigma $-field ($\vec{\pi}=0$) for different proper times (indicated in the figure).  The zero-temperature potential $U(\sigma,0)$ is also shown.\\
 \\
 \\
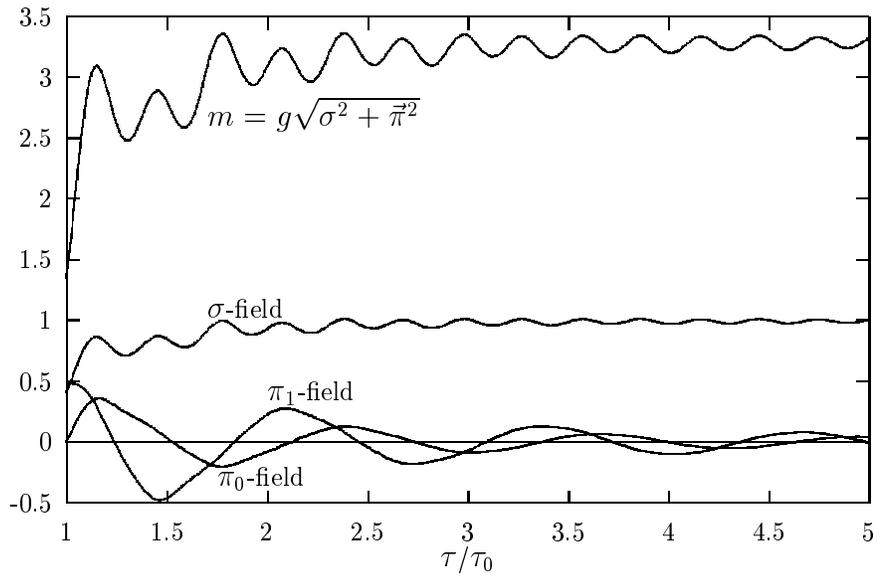
Figure 2: Meson fields and constituent quark mass in (units of $f_{\pi}$) as functions of proper time (in units of $\tau_{0}=7$ fm/c) calculated for the sigma model. The initial conditions at $\tau =\tau _{0}$ are: $\Phi =(38,0,45,0)$ MeV and $\frac{\partial \Phi }{\partial \pi}=(40,35,0,0)$ MeV/fm.\\
 \\
 \\
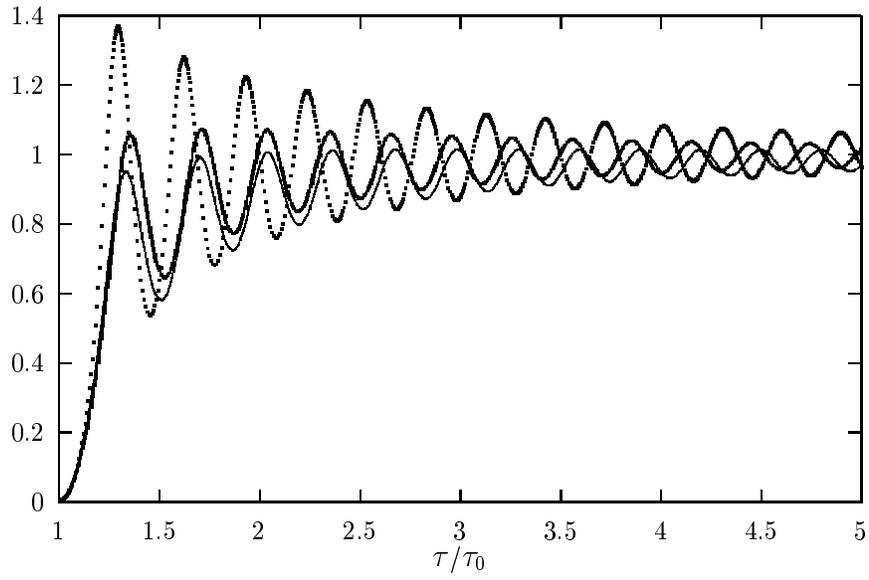
Figure 3: Proper time evolution of the $\sigma$-field calculated in the sigma model for three different scenarios. Dotted curve corresponds to the quench.
The other curves are calculated for scaling expansion with mesonic quasi-particles (solid line) and quark-antiquark quasi-particles (bold line). The initial conditions are all zero\\
 \\
 \\
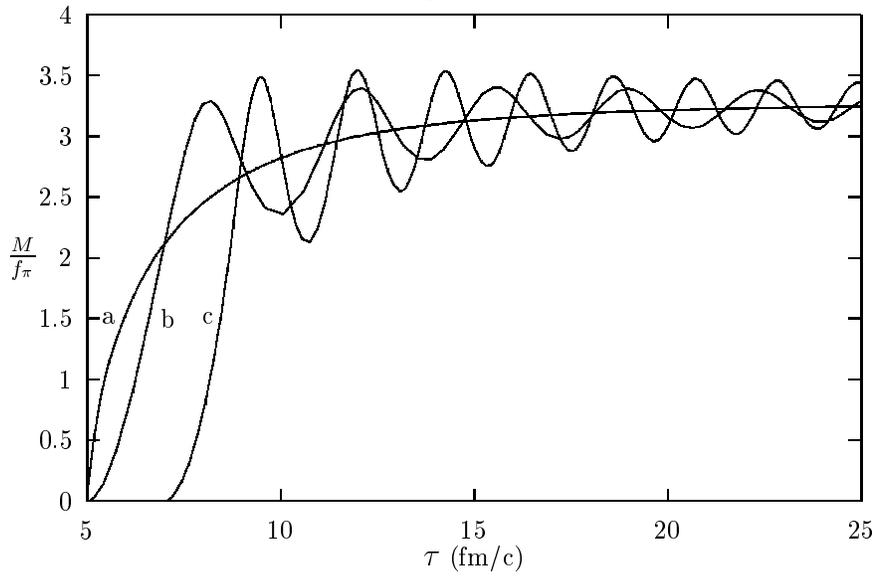
Figure 4: Constituent quark mass as a function of proper time calculated within the NJL model without (a) and with (b) derivative terms. For comparison the result for the linear sigma-model is also shown (c). Explicit symmetry breaking terms are included and zero initial conditions are imposed in cases (b) and (c).

\end{document}